\documentstyle[11pt,epsfig]{article}

\def\laq{~\raise 0.4ex\hbox{$<$}\kern -0.8em\lower 0.62
ex\hbox{$\sim$}~}
\def\gaq{~\raise 0.4ex\hbox{$>$}\kern -0.7em\lower 0.62
ex\hbox{$\sim$}~}

\def\beq{\begin{equation}}
\def\eeq{\end{equation}}
\def\bea{\begin{eqnarray}}
\def\eea{\end{eqnarray}}
\def\bean{\begin{eqnarray*}}
\def\eean{\end{eqnarray*}}

\def\vp{\varphi}
\def \vpb {{\overline {\vp}}}
\def \vpbp {{\dot{\vpb}}}

\def \pa {\partial}
\def \ra {\rightarrow}

\def \la {\lambda}
\def \ls {\lambda_{\rm s}}

\def \Ms {M_{\rm s}}
\def \Mp {M_{\rm P}}

\def \da {\delta}

\def \a {\alpha}
\def \ap {\alpha^{\prime}}

\def \ga {\gamma}

\def \da {\delta}

\def \om {\omega}

\begin{document}

\begin{titlepage}

\begin{flushright}
BA-TH/03-463\\
CERN-TH/2003-128\\
hep-th/0306113
\end{flushright}

\vspace*{1.8 cm}

\begin{center}

\huge
{Perturbations in a non-singular bouncing Universe}

\vspace{1cm}

\large{M. Gasperini$^{1}$, M. Giovannini$^2$ and G. Veneziano$^2$}

\normalsize
\vspace{.2in}

{\sl $^1$Dipartimento di Fisica, Universit\`a di Bari, \\ 
Via G. Amendola 173, 70126 Bari, Italy\\
and \\
Istituto Nazionale di Fisica Nucleare, Sezione di Bari\\
Via G. Amendola 173, 70126 Bari, Italy}

\vspace{.2in}

{\sl $^2$Theoretical Physics Division, CERN, \\
CH-1211 Geneva 23, Switzerland }

\vspace*{1.5cm}

\begin{abstract}
We complement the low-energy gravi-dilaton 
effective action of string theory  with a non-local, general-covariant
dilaton potential, and  obtain homogeneous solutions describing 
a non-singular (bouncing-curvature) cosmology.  We then compute, both
analytically and numerically, the spectrum of  amplified scalar and
tensor perturbations, and draw some  general lessons on
how to extract observable consequences from  pre-big bang and
ekpyrotic scenarios. 

\end{abstract}
\end{center}

\end{titlepage}
\newpage

In recent years string/M-theory has inspired new cosmological 
scenarios in which a long period of accelerated (growing-curvature)
evolution, emerging from an almost trivial initial geometry, turns into a
standard (decreasing-curvature)  FRW-type cosmology, after going smoothly
through a big bang-like event. There are by now several variations on
this pre-big  bang theme. Besides the original pre-big bang (PBB)
scenario \cite{1,2,3}, based on the duality symmetries of string
cosmology,  new models incorporating brane and M-theory ideas
have been proposed under the generic name of ekpyrotic (EKP)  scenarios
\cite{4}.
While different proposals differ in the way the scale factor behaves 
during the growing-curvature phase, they all share the  feature
of describing a bounce in $|H|$ (the absolute value
of the Hubble parameter) or, in more geometrical terms, in the  space-time
curvature. The latter always starts vanishingly
small, grows to a maximum at the would-be big bang, and then decreases again in
the FRW phase. A common theoretical challenge to all these models is
that of being able to describe the transition between the two regimes.

At a more phenomenological level, instead, the challenge is to compute,
in a reliable way, the final spectrum of amplified quantum fluctuations
to be compared with present data on CMB radiation and large-scale
structure.
In the PBB case it was admitted early on \cite{5,6} that adiabatic- 
curvature perturbations had too large a spectral index to be of any
relevance at cosmologically interesting scales (while being possibly
important for gravitational waves searches \cite{7,8}). Isocurvature
perturbations (related to the Kalb--Ramond two-form)  can instead  be
produced with an interestingly flat spectrum \cite{9}, but have to be
converted into adiabatic-curvature perturbations through the so-called
curvaton mechanism \cite{10} before they can provide a viable
scenario for large scale-anisotropies \cite{11}.
Proponents of the ekpyrotic scenario, while agreeing with the PBB 
result of a steep spectrum of tensor perturbations, have also
repeatedly claimed \cite{4,12} to obtain ``naturally" an almost
scale-invariant spectrum of adiabatic-curvature perturbations, very
much as in ordinary models of slow-roll inflation. These claims have
generated a lot of heated discussion (see
for instance \cite{13}), with many arguments given  in favour or against
the phenomenological viability of  EKP scenarios  in the absence of a
curvaton's help. The reasons for the disagreement  can be ultimately
traced back to the fact that the curvature bounce (hereafter simply
referred to as the bounce) is put in by hand, rather than being derived
from an underlying action. This leaves different authors to make
different assumptions on how to smoothly connect  perturbations across
the bounce itself, and this often results in completely different physical
predictions.

Our aim is to present a class of models where a regular 
bounce is derived from the field equations of a 
general-covariant, albeit non-local,  action, 
and where perturbations can be
studied from beginning to end.  We are thus able to test, at least within
the model, the assumptions made by different groups about how
different perturbations should or should not behave across the bounce.
In this letter we only consider a simple 
gravi-dilaton model and present  the main results,
working in the string frame and skipping their detailed derivation. 
In a longer, forthcoming paper \cite{16} we shall give
full details of the calculations in both the string and Einstein 
frames (with identical physical conclusions), and generalize the class of models 
to include a fluid component.
We first  present the action, the
background-field equations, and a class of smooth, bouncing solutions.
We then discuss, successively, the amplification of tensor and scalar 
perturbations from an initial spectrum of vacuum
quantum fluctuations, and  compare analytical and
numerical results. We finally summarize our results and draw some conclusions.

We recall that the homogeneous cosmological solutions derived from the tree-level, low-energy
string effective action exhibit in general a curvature singularity,
disconnecting the pre-big bang branch from the post-big bang one \cite{BV}. 
Such a singularity is expected to
be removed by higher-loops {\em and} higher-curvature $\ap$
corrections (see for instance \cite{3}), but it is known \cite{14} that non-singular
solutions can be explicitly obtained already at low curvatures \cite{2}
in the presence of an appropriate {\em non-local} effective potential.
An example of such a possibility can be illustrated by  using a
$(d+1)$-dimensional, general covariant  gravi-dilaton
effective action, which,  in the string frame, reads: 
\beq
S= -{1\over 2\ls^{d-1}} \int d^{d+1}x \sqrt{|g|}~
e^{-\vp}\left[R+ \left(\nabla \vp\right)^2+V(\vpb)\right]
\label{action}
\eeq
(metric conventions: $+--- \dots$).  Here $\ls= \sqrt{\ap} = \Ms^{-1}$ is the string
length scale, and the potential $V(\vpb(x))$, a local function of $\vpb$, is instead a
non-local function (yet a scalar under general coordinate transformations)
of the dilaton owing to the definitions:
\beq 
V= V(e^{-\vpb}),  ~~~~~~
e^{-\vpb(x)} =    \int {d^{d+1}y \over \ls^d}\sqrt{|g(y)|}~ e^{-\vp(y)}
\sqrt{\pa_\mu \vp (y) \pa^\mu \vp(y)}
~\delta\left(\vp (x) -   \vp (y)\right) .
\label{21}
\eeq
Note that $e^{\vpb}$ plays the role of a ``reduced" coupling constant from $d+1$ to $0+1$ space-time dimensions.
 We may thus expect $V(e^{-\vpb})$ to go like some inverse power of its argument as this becomes large (i.e. in perturbation theory). 
We shall discuss in \cite{16} how such a non-local potential may be
induced by loop corrections in (higher-dimensional) manifolds
with compact spatial sections; here, we take  this simply as a  toy model that avoids the singularity, 
while staying all the time at low energy/curvature.
For a background manifold isometric with respect to $d$ spatial
translations, it is known that  action (\ref{action}) leads to field equations that are
covariant under scale-factor duality \cite{1} and, in  the presence of the two-form background $B_{\mu\nu}$,  also under global $O(d,d)$ transformations \cite{14}. 

Variation of the above action with respect to $g_{\mu\nu}$ and
$\vp$, though somewhat unusual, is straightforward and leads to the following field equations  (in units of
$2\ls^{d-1}=1$):
\beq
G_{\mu\nu} + \nabla_{\mu} \nabla_{\nu} \vp + \frac{1}{2} g_{\mu\nu}  
\left[ (\nabla \vp)^2 - 2 \nabla^2 \vp - V\right] -  \frac{1}{2} e^{-\vp} 
\sqrt{(\partial \vp)^2} ~\ga_{\mu\nu} I_1 = 0,
\label{genback1}
\eeq
\beq
R + 2 \nabla^2 \vp - (\nabla \vp)^2 + V - {\pa V\over \pa{\bar{\vp}}} 
+  e^{-\vp}  \frac{\hat{\nabla}^2 \vp}{\sqrt{(\partial \vp)^2}}  I_1 - 
e^{-\vp}  V'   I_2 = 0,
\label{genback2}
\eeq
where (with a prime denoting differentiation with respect to the argument)
\begin{eqnarray}
&&
I_1 = {1\over \ls^d}\int d^{d + 1}y \sqrt{|g(y)|}~V'(e^{-\vpb(y)}) 
~\delta\left(\vp (x) -   \vp (y)\right) , 
\nonumber\\
&&
I_2 = {1\over \ls^d} \int d^{d + 1} y \sqrt{|g(y)|}~ 
\sqrt{\pa_\mu \vp (y) \pa^\mu \vp(y)}
~\delta'\left(\vp (x) -   \vp (y)\right) \; ,
\end{eqnarray} 
and we have also  introduced the induced metric and Laplacian:
\beq
\gamma_{\mu\nu} = g_{\mu\nu} - \frac{\partial_{\mu}\vp \partial_{\nu}\vp}{(\partial \vp)^2},  
~~~~~~~~~~  \hat{\nabla}^2 \vp = \ga_{\mu\nu} \nabla^\mu\nabla^\nu \vp.
\eeq
For a homogeneous, isotropic and spatially flat background, we can set 
$g_{00}=1$, $g_{ij}=-a^2(t)\da_{ij}$, $\vp=\vp(t)$, and  obtain $e^{-\vpb}= e^{-\vp}a^d$,
where we have absorbed into $\vp$ the dimensionless constant $-\ln 
(\int d^dy/\la_s^d)$, associated with the (finite) comoving spatial volume. 
In such a case, the time and space components of Eq.(\ref{genback1}), and the dilaton equation (\ref{genback2}), lead  to a set of equations already studied in \cite{14,2,5}:
\begin{eqnarray} 
&&
\dot{\overline{\vp}}^2 - d H^2 -V=0, ~~~~~~~~~~~~~~~~~~~~~
\dot H -H\vpbp=0,
\label{29}\\
&&
2 \ddot \vpb - \vpbp^2 -d H^2 +V -{\pa V\over \pa \vpb} =0,
\label{210}
\end{eqnarray}
where the  third (dilaton) equation follows from  the first two, {\em provided} $\vpbp\not=0$. 

As noticed in \cite{14}, the above set of equations admits, quite generically, non-singular solutions. To recall this point and give an explicit analytical  example we note that the two equations (\ref{29}) can be reduced to quadratures, i.e.  
\beq
H = m e^\vpb, ~~~~~~~~~
t= \int^\la d\la' \left[d+\la'^2 V(m\la')\right]^{-1/2} \;, 
\eeq
where $m$ is an integration constant, and $\la=m^{-1} e^{-\vpb}$. If the function appearing above between square
brackets has a simple zero, we obtain a regular bouncing solution \cite{14}. Furthermore, if $V\la^2 \ra 0$
at large $|\la|$ (corresponding to a potential generated beyond two loops), the asymptotic solutions approach the
usual vacuum solutions at large $|t|$. 

 Consider, in particular, the class of  potentials
\beq
V(m\la)= \la^{-2}\left( \left[\a -(m\la)^{-2n}\right]^{2-1/n} - d \right),
\label{212}
\eeq
parametrized by the dimensionless coefficients $\a$ and by the ``loop-counting" parameter $n$. For $\a >0$ and $n>0$, Eq. (\ref{212}) leads  to the general exact solution
\beq
H =m e^\vpb =m\left[\a \over 1+(\a mt)^{2n}   \right]^{1/2n}.
\label{213}
\eeq
As $|t| \ra \infty$, the Hubble parameter behaves like $H|t| \sim \a^{(1-2n)/2n}$, so that the minimal pre-big bang solutions dominated by the dilaton kinetic energy are only recovered for $ \a^{(2-1/n)}=d$ (see for instance \cite{3}). If this condition is not satisfied, the background may still be non-singular, but  the dilaton potential cannot be neglected, even  asymptotically.

 For the perturbation analysis of this paper it will be sufficient to use as a toy model the simple regular background  associated to the four-loop potential 
$V(\vpb)=-V_0e^{4 \vpb}$. In this case the general solution, in $d=3$ spatial dimensions, can be written as:
\beq
a(\tau) =  \biggl[ \tau + \sqrt{\tau^2 + 1}\biggr]^{1/\sqrt{3}},
~~~~~~~~~~
 \overline{\vp} = - \frac{1}{2} \ln{ ( 1 + \tau^2)} + \vp_{0},  
~~~~~~~~~~ \tau=t/t_0, 
\label{214}
\eeq
where $\vp_0$ is an integration constant and $t_0^{-1} = e^{{\vp}_0} \sqrt{V_0}$. The solution is thus characterized 
by two relevant parameters,  $t_0^{-1}$ and $e^{{\vp}_0}$, corresponding, up to some numerical factors, to the
Hubble parameter and string coupling at the bounce, respectively (without loss of generality we have set 
$a(0) =1$ with $t=0$, the time at which the bounce occurs).
We stress that the possibility of regular backgrounds is not limited to the class of potentials illustrated in Eq. (\ref{212}), and that regular bouncing solutions can also be obtained by adding to the action (\ref{action}) fluid matter sources, as already pointed out in \cite{2} and  illustrated in  \cite{16}.

The evolution equation of tensor (transverse and traceless) 
metric perturbations   can be obtained by perturbing 
to first order 
the $(i,j)$ component of Eq. (\ref{genback1}). For 
each polarization we obtain, in Fourier space,
\begin{equation}
\ddot{h}_{k} - \dot{\overline{\vp}} \dot{h}_k + \om ^2 h_k =0 \;  ,  
~~~~~~~~~~\om \equiv {k}/{a}.
\label{GW1}
\end{equation}
We shall concentrate our attention on the case of regular 
solutions approaching asymptotically  the well- 
known minimal gravi-dilaton model with negligible potential.
The gravitational wave spectra arising in this  case
 are characterized, in general,   by a steep  spectrum ($n_T >1$), up to
 logarithmic corrections. In fact the asymptotic 
solution of Eq. (\ref{GW1}) for large wavelengths can be written as
\begin{equation}
h_{k} = A_{k} + B_{k} \int_{t_{\rm ex}}^{t} 
\frac{e^{\vp}}{a^{3}} dt,
\label{outhor}
\end{equation}
where $t_{\rm ex} \sim \om^{-1}$ denotes the time at which the perturbation 
exits the horizon. Since in the case under consideration
 $e^{\vp}a^{-3} \sim  |t|^{-1}$ for $t \to -\infty$,
we must expect 
 a $\ln{\om t}$  growth of $h_{k}$,  leading,  ultimately, to
$\ln{k}$  corrections in the power spectrum.

For an accurate camparison of analytical and numerical 
results, we shall restrict our attention to the 
specific regular bouncing 
 solution (\ref{214}). In this case, the evolution 
equation for the canonical normal mode 
$\mu_{k} = a e^{-\vp/2} h_{k}$ has asymptotic solutions with 
normalization to an initial vacuum fluctuation spectrum
given in terms of Hankel functions of index zero. The application 
of the standard matching procedure leads to  a Bogoliubov 
coefficient (see below for its exact definition) given by 
\begin{equation}
|\beta_{k}|^2 =\epsilon_{1} + \epsilon_{2}\ln^2({k_1/k}),
\label{exptensor}
\end{equation}
where $\epsilon_{1}$ and $\epsilon_2$ are numerical factors 
of order $1$,  and $k_1$ is a characteristic 
momentum scale. The associated power spectrum is 
\begin{equation}
|\delta h_{k}|^2 = k^3 |h_{k}|^2 \sim\left(H(0)\over \Mp(0)\right)^2
 \left(k\over k_1\right)^2 \ln^2\left(k\over k_1\right),
\label{dhnorm}
\end{equation}
where the normalization $H(0)/M_{P}(0) \sim (\lambda_{s} /t_{0})~
 e^{\vp(0)/2}$  is controlled by the ratio between the 
curvature scale and the Planck mass at the bounce $t =0$. 

For the numerical calculation, it is convenient 
to work with the rescaled variable $\hat{h} = e^{-\vp/2} a^{3/2} h =
\sqrt{a} \mu $. In order to compute the amplification, we impose 
for $t \to -\infty$ the quantum mechanical initial conditions 
\begin{equation}
\mu_{k} = \frac{1}{\sqrt{2 k} } e^{- i k \eta}~~~~~\rightarrow~~~~~
\hat{h}_{k} = \frac{1}{\sqrt{2 \omega} } e^{ - i \int \omega dt},
\end{equation}
where we recall that we are using units in which
$\Ms^2 = 2$. The action of the gravi-dilaton background 
will  produce a mixing in the positive and negative frequency modes so
that,  for $t \to +\infty$, the solution  can be parametrized as 
\begin{equation}
\hat{h}_{k} =\frac{1}{\sqrt{2 \omega}} \left(\alpha_{k} e^{- i\int 
\omega dt} + \beta_{k} e^{i \int \omega dt}\right).
\label{18}
\end{equation}
In a second quantized approach $|\beta_{k}|^2$ is nothing but the 
number of produced gravitons of given momentum $k$. The energy 
density of the produced gravitons will then be, up to numerical factors, 
$\int d \ln{k}~ k^4 |\beta_{k}|^2$. Thus, using the 
theoretical expectations given above, this 
corresponds to  an  
energy  spectrum going as $k^4$ (up to logarithmic corrections).
\begin{figure}[!ht]
\centerline{\epsfxsize = 11cm  \epsffile{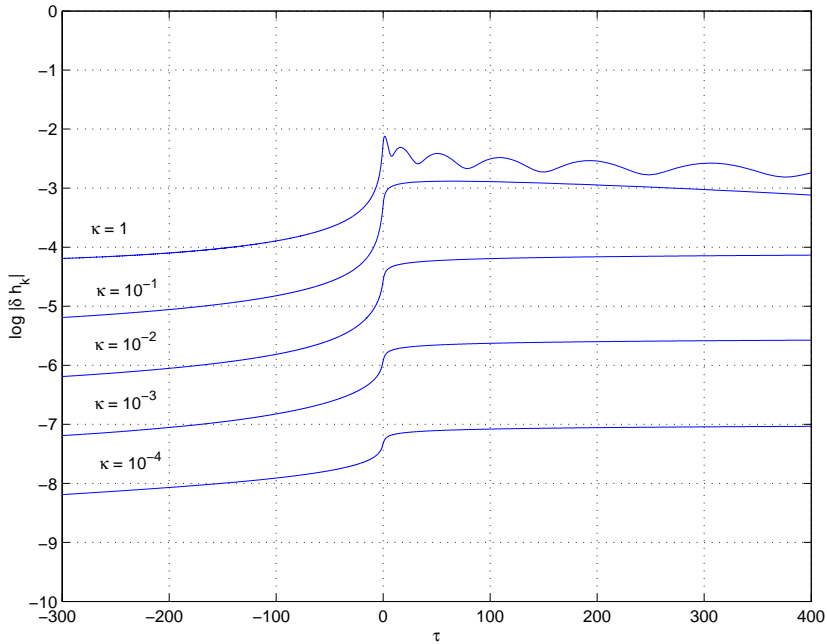}}
\vskip 3mm
\caption[a]{Behaviour of the tensor power  spectrum $|\delta h_{k}|$, for different values of $\kappa = k t_0$, 
in the region around  the bounce. In the present and in the following figures the $\log$ denotes the 
logarithm in basis ten.}
\label{FIGURE1} 
\end{figure}
The numerical integration in terms shows that 
both  $\hat{h}$ and $h$ pass regularly through the bounce; this 
is  illustrated in Fig. \ref{FIGURE1}, 
where we report the tensor power spectrum, $|\delta h_{k}|$, 
in the region around $t=0$, clearly in qualitative agreement with Eq. (\ref{dhnorm}).The  absolute normalization 
of the tensor power spectrum has been fixed by imposing $H(0)/M_{P}(0)=10^{-2}$.

\begin{figure}
\centerline{\epsfxsize = 11cm  \epsffile{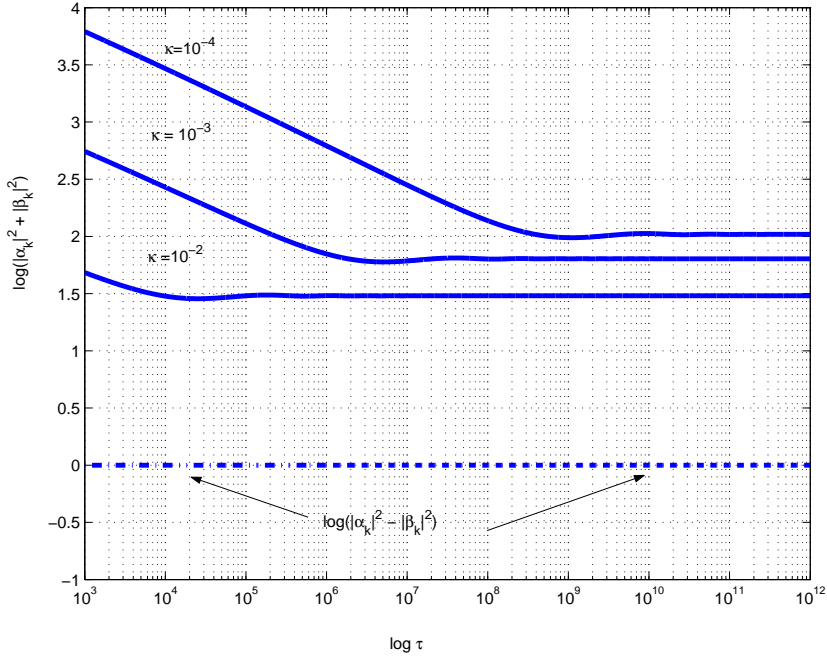}}
\vskip 3mm
\caption[a]{Asymptotic behaviours  of the mixing coefficients given 
in Eqs. (\ref{sum}) and (\ref{difference}) for different values of 
$\kappa =k t_{0}$.}
\label{FIGURE2} 
\end{figure}

The mixing 
coefficients of Eq. (\ref{18}) can be obtained in terms 
of the asymptotic  ($\tau \to +\infty$) values of the real and imaginary parts of $\hat{h}$,  via the expressions:
\begin{eqnarray}
&& |\alpha_{k}|^2 + |\beta_{k}|^2 = \left( \omega |\hat{h}_{k}|^2  
+ \frac{1}{\omega}\left| \dot{\hat{h}}_{k} - \frac{H}{2} \hat{h}_{k}\right|^2
 \right),
\label{sum}\\
&& |\alpha_{k}|^2 - |\beta_{k}|^2= i \biggl( \dot{\hat{h}}_{k} \hat{h}_{k}^{\ast} 
-{\dot{\hat{h}}}^{\ast}_{k} \hat{h}_{k}\biggr). 
\label{difference}
\end{eqnarray}
We expect Eq. (\ref{difference}) to be identically 
equal to $1$ thanks to the conservation of the Wronskian.  On the other hand, Eq. (\ref{sum}) 
should  approach a ($k$-dependent) constant only at late 
enough times.
These two expectations are perfectly fulfilled, as  illustrated in Fig.
\ref{FIGURE2}, which thus represents 
a highly non-trivial consistency check of our numerical
procedure.

\begin{figure}
\centerline{\epsfxsize = 11cm  \epsffile{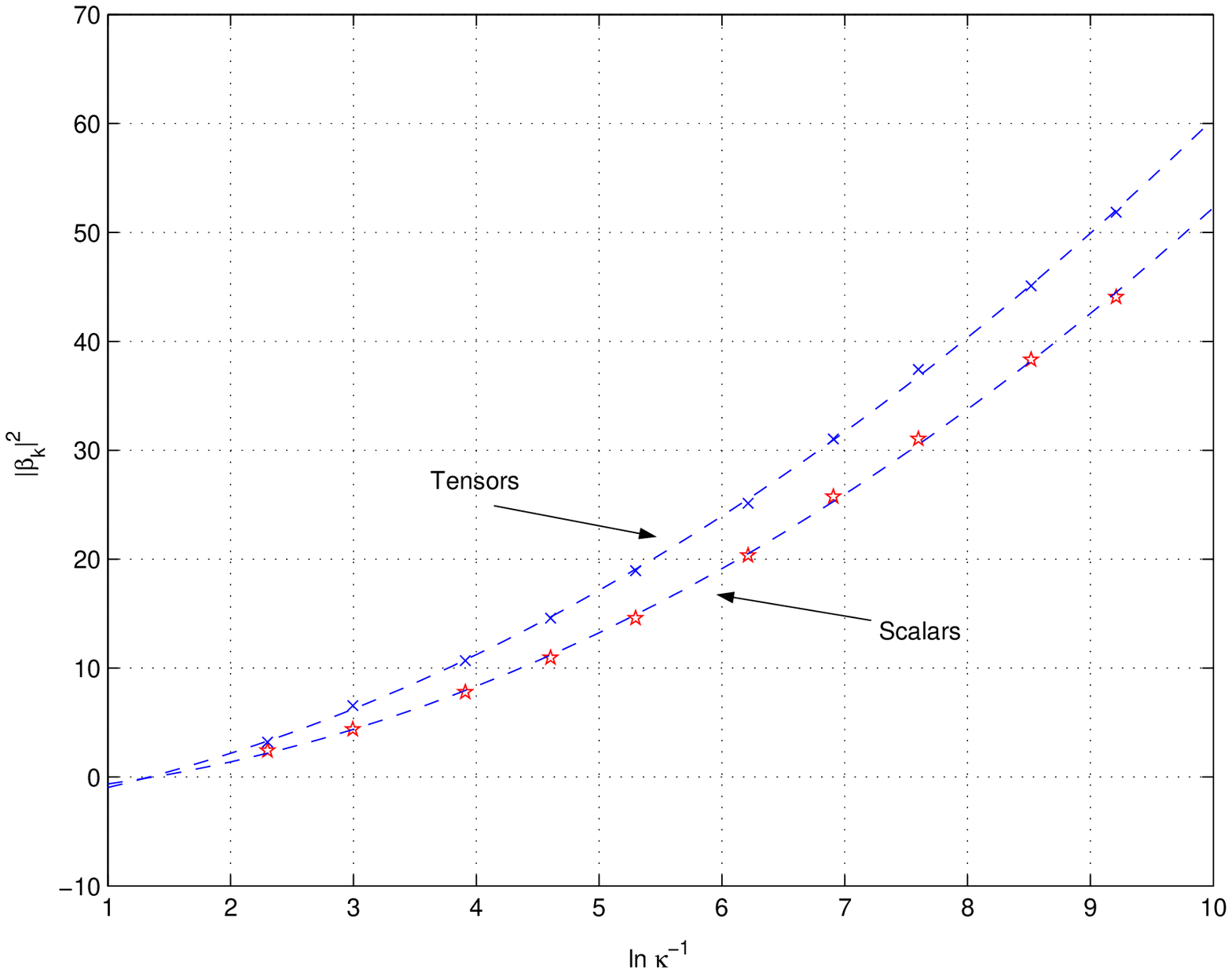}}
\vskip 3mm
\caption[a]{ The numerically determined values of  $|\beta_k|^2$  are fitted to theoretical expectations for tensor (crosses) and scalar (stars) perturbations.}
\label{FIGURE3} 
\end{figure}

Clearly,  smaller $k$ re-enter the horizon (and thus reach their asymptotic value) later.
The asymptotic value gives, for each $k$, the sought  after Bogoliubov  coefficient. The results are plotted in Fig.
\ref{FIGURE3} (crosses) and fitted to the theoretical expectation (\ref{exptensor}).
The fit is very good and the parameters $\epsilon_{1} , \epsilon_{2}, k_1$ are determined according to:
\begin{equation}
|\beta_{k}|^2 = 0.46 \ln^2(k/k_1) -4.84, ~~~~~~~~~~~~~ k_1 = 
6.68/t_{0}.
\end{equation}
We have not yet attempted to make a theoretical estimate of  $\epsilon_{2}$.

 Unlike tensor perturbations, scalar fluctuations are, in general, neither gauge- nor frame-independent.
Gauge-invariant perturbations can be defined, as usual, in each frame. Among these,  the spatial curvature perturbation  on uniform dilaton 
hypersurfaces ${\cal R}$, and the Bardeen potentials $\Psi$ and $\Phi$ are   particularly important for CMB
phenomenology. Denoting by $\delta \varphi = \chi$ the dilaton perturbation, and introducing  scalar metric perturbations in conformal time via the standard expressions 
\begin{equation} 
 \delta g_{00} = 2 a^2 \phi,~~~~~~~\delta g_{i j} 
= 2 a^2 ( \psi \delta_{ij} - \partial_{i}\partial_{j } E), ~~~~~~~ \delta g_{i 0} = a^2 \partial_{i} B  \; ,
\label{linepert}
\end{equation}
we may express  ${\cal R}$,  $\Psi$ and $\Phi$ as:
\begin{equation}
{\cal R} = - \psi - \frac{{\cal H} }{\varphi'} \chi,~~~
 \Psi = \psi + {\cal H} (E' - B),  ~~~ \Phi = \phi - {\cal H} (E' - B) - (E' - B)',
\end{equation}
where primes denote derivatives with respect to conformal time and 
${\cal H}=a'/a$.
Introducing the Einstein frame, with scale factor $a_E=a e^{-\vp/2}$, 
we have $\phi_E = \phi_s - \frac{1}{2} \chi$,  $\psi_E = \psi_s  + \frac{1}{2} \chi$ (while $B$ and $E$ coincide in the two frames), and  we can easily prove that
${\cal R}$ and  $\Psi +  \Phi$ are frame-independent,  while $(\Psi -  \Phi)_s = (\Psi -  \Phi)_E - \chi + \vp' E'$.

We shall work in the convenient 
 uniform dilaton gauge $\chi = 0$, and also set $B=0$, so that
${\cal R} = - \psi$ and $\Psi_E = \psi + {\cal H}_{E} E'$, where
${\cal H}_{E}= {\cal H} - {\vp'}/{2}$. 
This gauge is particularly useful since all the perturbations variables in (\ref{linepert}) coincide in the two frames. Another 
advantage is that, by perturbing the background equation (\ref{genback1}), we find in this gauge $\delta \gamma_{\mu}^{\nu}=0$,
$\delta e^{-\vpb}=0$, $\delta I_1 =0$.
The perturbations of the background equations 
thus simply lead to the 
following set of evolution equations:
\begin{eqnarray}
&& 3 (\vp' - 2 {\cal H}) \psi' + ( 6 {\cal H} \vp' - {\vp'}^2 - 6 {\cal H}^2) 
\phi + 2 \nabla^2 \psi - (\vp' - 2 {\cal H}) \nabla^2 E' =0, ~~~~~
\label{00P1}\\
&& (\vp' - 2 {\cal H}) \phi = 2 \psi',~~~~~~~~~~~~~~~~~~~~~~~~~~
~~~~~~~~~~~~~~
\label{i0P1}\\
&& E'' + (2{\cal H} - \vp') E' + \psi - \phi =0, ~~~~~~~~~~~~~~~~~~~
~~~~~~~~~
\label{ineqjP1}
\end{eqnarray}
following,  respectively, from the 
$(00)$, $(0i)$ and  $(i\neq j)$ components of the  perturbed 
Eq. (\ref{genback1}).  Note that the last equation, upon using ${\cal H}_{E}= {\cal H} - {\vp'}/{2}$,  is nothing but $(\Psi -  \Phi)_E =0$, and that
the additional equations obtained from the $(i=j)$ component 
and from the dilaton equation (\ref{genback2}) are redundant.

We may combine the above system to get the interesting equation 
\begin{equation}
\psi' = \frac{( \vp' -2 {\cal H})}{{\varphi'}^2} \nabla^2 
 \left[ ( \vp' -2 {\cal H})E' - 2 \psi \right],
\label{pspr}
\end{equation}
which  turns out 
to be  exactly the standard evolution equation for
${\cal R}$:
\begin{equation}
{\cal R}' = - \frac{ 4 {\cal H}_{e}}{{\vp'}^2} \nabla^2  \Psi_E.
\label{R}
\end{equation}
Although this equation  is {\em formally } the same 
as the standard equation  of ordinary
cosmological perturbation theory, the relation between
$\varphi'^2$ to $({\cal H}_{E}^2 - {\cal H}_{E}')$ 
is different in our case. 
Indeed, it can be easily  checked, by transforming the background equations (\ref{29}), (\ref{210}) to the Einstein frame, that:
\begin{equation}
4 ({\cal H}_{E}' - {\cal H}_{E}^2) + {\vp '}^2 + 
\frac{\partial V}{\partial \overline{\vp}} a_{e}^2 e^{\vp}=0.
\end{equation}
This shows that the usual identity 
$ {\vp '}^2 \propto  ({\cal H}' - {\cal H}^2)$ is not satisfied
in our case. And indeed, in our 
class of bouncing solutions
$\varphi'$, unlike ${\cal H}_E' - {\cal H}_E^2$,  {\em never } vanishes.
As a consequence, the pre-factor on the right-hand side 
of Eq. (\ref{R}) does not diverge during the bounce; we can thus infer, already at this 
stage, that ${\cal R}$ is likely to be conserved on super-horizon scales.

Let us now turn to the specific study 
of the background model already analyzed in the case of
the tensor perturbations.
On the basis of simple estimates based 
on the asymptotic behaviour of the solutions prior to the bounce and after, 
we expect the amplification coefficients for the 
scalar and the  tensor modes of the geometry to be very similar, both
qualitatively and quantitatively. Indeed, after some lengthy 
but trivial algebra, the following decoupled evolution equation for 
${\cal R}$ 
can be obtained from Eqs.(\ref{00P1})--(\ref{ineqjP1}):
\begin{equation}
{\cal R}_{k}'' + 2 \frac{z'}{z} {\cal R}_{k}' + k^2 C_{s}^2 {\cal R}_{k} =0,
\end{equation}
where  
\begin{equation}
z = \frac{\varphi'}{{\cal H}_e} a e^{-\vp/2},~~~~~
~~~~
C_{s}^2= \left( 1 + \frac{\partial V}{\partial \vpb} \frac{a^2}{{\vp'}^2} 
\right).
\end{equation} 
On the other hand, in order to enforce the correct quantum normalization, we note 
that ${\cal R}$ is related to the canonical normal mode of scalar perturbations, 
$v_{k} = z {\cal R}_{k}$. For large $|t|$, $V\to 0$, $z \propto 
a e^{- \vp/2}$ and we are led exactly to the same 
equations as for tensor perturbations, the same asymptotic solutions, and thus 
the same spectrum for $\delta R_{k}$.

These expectations will now be checked  numerically. 
We first notice that it is impossible to follow the evolution of perturbations directly through the canonical variable
$v$ (or the other often used variable $u$), since these become singular at the (Einstein-frame) bounce. This is not a problem, however, since $v$ is only needed at very early or very late times, for normalization purposes.
We can instead integrate directly a first-order system of differential equations,  
involving $\psi =- \cal{R}$ and $E'$, which is completely regular throughout. Going over to (string-frame) cosmic time, and using 
the constraint (\ref{i0P1}) into Eqs. (\ref{00P1}) and 
(\ref{ineqjP1}), we can eliminate $\phi$. We then obtain:
\begin{equation}
 \dot{\psi}_{k} = A_{k}(t) \psi_{k} + B_{k}(t) {\cal E}_{k}, ~~~~~~~~~
 \dot{{\cal E}}_{k} = C_{k}(t) \psi_{k} + D_{k}(t) {\cal E}_{k},
\label{seconeq}
\end{equation}
where ${\cal E}_{k}= a^2 \dot{E}_{k}$, and 
\begin{eqnarray}
A_{k}(t) = \frac{2 (\dot{\vp} - 2  H)}{\dot{\vp} }\omega^2, ~~ &&~~ B_{k}(t) =-\biggl(\frac{\dot{\vp} - 2  H}{\dot{\vp}}\biggr)^2 
\omega^2,
\nonumber\\
C_{k}(t) = \frac{4\omega^2}{ \dot{\vp}^2} -1 ,~~ &&~~ 
D_{k}(t) = \dot{\varphi} -  H  - A_{k}(t).
\label{ABCD}
\end{eqnarray}
 This system can be solved by imposing
quantum mechanical initial conditions for  the fluctuations of the
curvature. Using $v_{k} = z {\cal R}_{k}$ we expect, 
asymptotically 
\begin{equation}
{\cal R}_{k} \sim \frac{1}{w} \frac{1}{\sqrt{2 \omega} } \biggl( \alpha_{k} e^{- i \int \omega dt} + 
\beta_{k} e^{ i \int \omega dt} \biggr),
\end{equation}
where $w= \sqrt{a} z$.

\begin{figure}
\centerline{\epsfxsize = 11cm  \epsffile{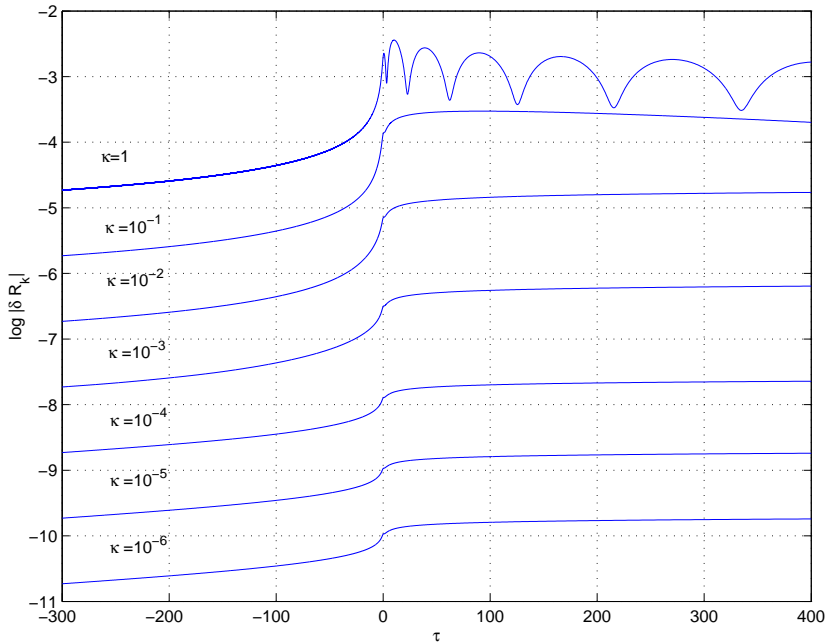}}
\vskip 3mm
\caption[a]{Behaviour of the power spectrum 
of ${\cal R}$, for different values of $\kappa = k t_{0}$.}
\label{FIGURE4} 
\end{figure}

Following the same lines the discussion as were developed in the case of  the
tensor modes of the  geometry, we can obtain, in the asymptotic region
after the bounce,   the appropriate combinations that  determine
the mixing coefficients: 
\begin{eqnarray}
|\alpha_{k}|^2 - |\beta_{k}|^2 &=&  i w^2 \left(  {\cal R}_{k}^* \dot{\cal R}_{k} - {\cal R}_{k} \dot{\cal R}_{k}^* \right) ,
\nonumber\\
 |\alpha_{k}|^2 + |\beta_{k}|^2 &= & w^2 \left( \omega  |{\cal R}_{k}|^2 
+\frac{1}{\omega}  |Y {\cal R}_{k} +\dot{\cal R}_{k}|^2\right)\; , 
\label{MI}
\end{eqnarray}
where $Y = \dot z/z$. 
The numerically computed  power spectrum of ${\cal R}$ is reported in Fig. \ref{FIGURE4}. 
In full analogy with the procedure discussed before, we have also  plotted the quantities appearing 
in Eqs. (\ref{MI}), checking that the first one gives identically $1$, and that the second approaches a limiting value at late times.
 We can thus extract the corresponding value of $|\beta_{k}|^2$  and fit it to the theoretical
expectation.
The results of this analysis  
are  reported in Fig. \ref{FIGURE3},  together with the already discussed results for the 
tensors. The fit for the scalar case (starred points) gives:
\begin{equation}
|\beta_{k}|^2 = 0.46 \ln^2(k/k_{2}) - 2.22,~~~~~~~~~~~~~~k_{2} 
= 2.2 /t_{0}.
\end{equation}
This formula is very similar to the one obtained in 
the case of tensors perturbations.

\begin{figure}
\centerline{\epsfxsize = 10cm  \epsffile{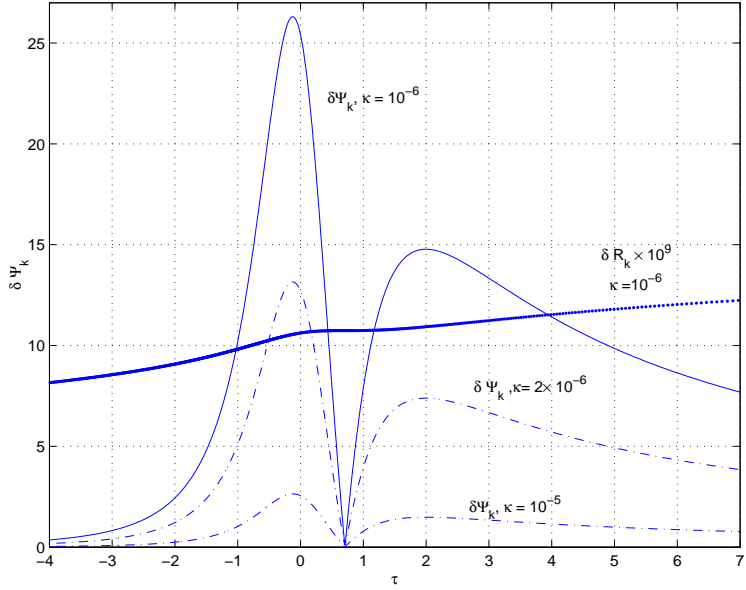}}
\vskip 3mm
\caption[a]{Time evolution of the power spectrum of the 
(Einstein-frame) Bardeen potential  $\Psi_E$ (thin and dashed curves).
 The bold curve represents the behaviour of an (appropriately rescaled) $\cal{R}$ mode.} 
\label{FIGURE5} 
\end{figure}

We now turn our attention to the behaviour of the (Einstein-frame) Bardeen potential. 
Its time evolution is shown in Fig. \ref{FIGURE5}, where it is clearly visible that  $\Psi$ 
becomes much larger than one for small $\kappa \eta$,  signalling a breakdown of perturbation theory in the 
longitudinal gauge (a point already emphasized by the first paper quoted in Ref. \cite{13}). 
The two Bardeen potentials in the string frame, though unequal, exhibit similar pathological behaviours.
We also see that, although all variables are regular across the bounce, the values of $\Psi$ and $\Psi'$, unlike those of $\cal{R}$,
change drastically across the bounce. Thus,
assuming continuity of $\Psi$ and of its derivative in simplistic models for the bounce is very dangerous.

Let us finally turn to another popular variable, the spatial-curvature perturbation on constant-density hypersurfaces, $\zeta$. 
It can be shown from its definition that it is not frame-independent, i.e. that
\begin{equation}
\zeta_s = - \psi_s  - \frac{{\cal H}_{s} }{\rho'_s} {\delta \rho_s}\not=
\zeta_E = - \psi_E  - \frac{{\cal H}_{E} }{\rho'_E} {\delta \rho_E}. 
\end{equation}
In any case, in our gauge, $\zeta_E$ and  $\zeta_s$  differ by $\cal{R}$ by terms proportional to $\phi$.
 The latter, however, is very suppressed at large scales (cf. Eqs. (\ref{i0P1}) and (\ref{pspr})). We conclude that
also $\zeta_s$ and $\zeta_E$ are well-behaved across the bounce. The fact that the power spectra of  both $\zeta_E$,  $\zeta_s$ and $\cal{R}$
stay small across the bounce implies that perturbation theory remains valid at all times in our gauge,  provided
$H(0)/M_P(0) < 1$.

We now summarize our results and draw some conclusions:
\begin{itemize}
\item Having defined a non-singular  bouncing cosmology, we are able to
follow the behaviour of the various perturbation variables from
beginning to end. \item Tensor perturbations behave as expected, with
$h$ becoming constant on superhorizon scales. Its power spectrum can
be computed by studying the associated canonical variable $\mu$ that
turns out to be also regular throughout the bounce. The numerically
computed power spectrum is in perfect agreement with the analytic 
expectations (including log corrections). 
\item For scalar perturbations, the Bardeen
potential $\Psi$,  the curvature perturbation on uniform dilaton
hypersurfaces $\cal{R}$, and the curvature perturbation on constant
 density hypersurfaces $\zeta$, all  go smoothly through the
bounce.  
\item Both $\cal{R}$ and $\zeta$ stay constant on superhorizon
scales, in agreement with general arguments \cite{w}, while $\Psi$ does not.
\item Provided that the ratio of the Hubble parameter to the effective Planck
mass  at the bounce ($t=0$) is small,  $H(0)/M_P(0) < 1$,  $\cal{R}$ and
$\zeta$ remain sufficiently small {\it at all scales} for perturbation
theory to remain valid at all times in the comoving gauge.
\item By
contrast, $\Psi$ becomes so large, at large scales, near the bounce that
perturbation theory breaks down in the longitudinal gauge.
\item The canonical variable $v$ of scalar perturbations 
(as well as another often discussed
variable, $u$) exhibits singularities at the bounce. This is not a problem
since the only use of $v$ is that of giving the initial normalization of the
fluctuations and the final Bogolubov coefficient, and $v$ is well behaved
at sufficiently early or late times.
\item We are thus able to compute
numerically the scalar perturbation power spectrum. We find that it is
very similar to that of tensor perturbations and in agreement with the
analytic results that follow by assuming a smooth behaviour of 
$\cal{R}$ (or $\zeta$) through the bounce (even when the background
itself has discontinuous derivatives). 
\end{itemize} 
In conclusion, at
least within the class of models considered in this paper, the procedure
used to argue  that, in bouncing Universes, the spectrum of adiabatic
scalar perturbations can be much flatter than the one 
of tensor perturbations, appears to be unjustified.

We are grateful to Valerio Bozza for help during the early stages of 
this work.

\end{document}